\documentclass[12pt]{iopart}
\usepackage{graphicx}

\def\bea{\begin{eqnarray}}
\def\eea{\end{eqnarray}}
\def\f{\frac}

\def\be{\beta}
\def\D{\Delta}

\def\r{\rho}
\def\a{\alpha}
\def\s{\sigma}
\def\kb{k_B}
\def\la{\langle}
\def\ra{\rangle}
\def\nn{\nonumber}

\def\bv{{\bf v}}

\def\br{{\bf r}}

\def\S{\Sigma}

\def\d{\delta}
\def\p{\partial}

\def\a{\alpha}
\def\d{\delta}
\def\p{\partial} 
\def\nn{\nonumber}
\def\r{\rho}

\def\rv{\vec{r}}

\def\la{\langle}
\def\ra{\rangle}

\def\f{\frac}

\begin{document}
\letter{Heat conduction through a trapped solid: effect of structural changes on thermal conductance}
\author{Debasish Chaudhuri$^1$, Abhishek Chaudhuri$^2$ and Surajit Sengupta$^3$}
\address{ $^1$Max Planck Institute for the Physics of Complex Systems,
N{\"o}thnitzer Str. 38, 01187 Dresden, Germany.}
\address{$^2$ Raman Research Institute,
Bangalore - 560080, India.}
\address{$^3$  Satyendra Nath Bose National Centre for Basic Sciences -
Block-JD, Sector-III, Salt Lake,
Calcutta - 700098.}
\eads{\mailto{debc@mpipks-dresden.mpg.de},\mailto{abhishek@rri.res.in},\mailto{surajit@bose.res.in}}

\begin{abstract}
We study the conduction of heat across a narrow solid strip trapped by
an external potential and in contact with its own liquid. Structural 
changes, consisting of addition and deletion of crystal layers in the 
trapped solid, are produced by altering the depth of the confining 
potential. Nonequilibrium molecular dynamics simulations and, wherever 
possible, simple analytical calculations are used to obtain the thermal 
resistance in the liquid, solid and interfacial regions (Kapitza or 
contact resistance). We show that these layering transitions are 
accompanied by sharp jumps in the contact thermal resistance. 
Dislocations, if present, are shown to increase
the thermal resistance of the strip drastically.
\end{abstract}
\submitto{\JPCM}
\maketitle

\section{Introduction}
The transport of heat through small and low dimensional
systems has enormous significance in the context of designing useful
nano-structures \cite{cahill}. 
Recently, it was shown\cite{abhi} that a narrow solid strip trapped by 
an external potential\cite{trap,phillips,grier,doyle-1} and 
surrounded by its own fluid relieves mechanical stress via 
the ejection or absorption of single solid layers\cite{myfail,ricci,degennes} 
to and from the fluid. 
The trapping potential introduces large energy barriers for interfacial  
capillary fluctuations thereby forcing the 
solid-liquid interfaces on either side of the solid region to 
remain flat. The small size of the solid also inhibits the creation of
defects since the associated inhomogeneous elastic displacement fields 
need to relax to zero quickly at the boundaries, making the elastic 
energy cost for producing equilibrium defects prohibitively large. Therefore 
the only energetically favourable fluctuations are those that  
involve transfer of {\em complete} layers which cause at most a 
homogeneous strain in the solid\cite{abhi}. 
Such layering transitions were shown to affect the sound absorption 
properties of the trapped solid in rather interesting ways\cite{abhi}.
What effect, if any, do such layering transitions have 
for the transfer of heat?  

Heat transport across model liquid-solid
interfaces has been studied in three dimensions for particles with Lennard
Jones interactions in Ref.\cite{barrat} along the liquid-solid coexistence 
line. The dependence of the Kapitza (interfacial) 
resistance\cite{kapitza1,kapitza}
on the wetting properties of the equilibrum interface was the focus of 
this study. It was shown that  a larger density jump at the
interface causes higher interfacial thermal resistance.
In this Letter, we use a nonequilibrium molecular dynamics simulation to 
investigate heat conduction through a trapped solid (in two dimensions) 
as it undergoes layering  
transitions as a response to changes in the depth of the trapping potential. 
Apart from the layering transition reported in Ref.\cite{abhi} we find another
mode of structural readjustment viz. increase in the number of atoms 
in the lattice planes parallel to the interface by the spontaneous 
generation and annihilation of dislocation pairs. The heat conductance,
in our study, 
shows strong signatures of both of these structural transformations of the 
trapped solid.

\section{System and simulations}
\begin{figure}[t]
\begin{center}
\includegraphics[width=5.cm]{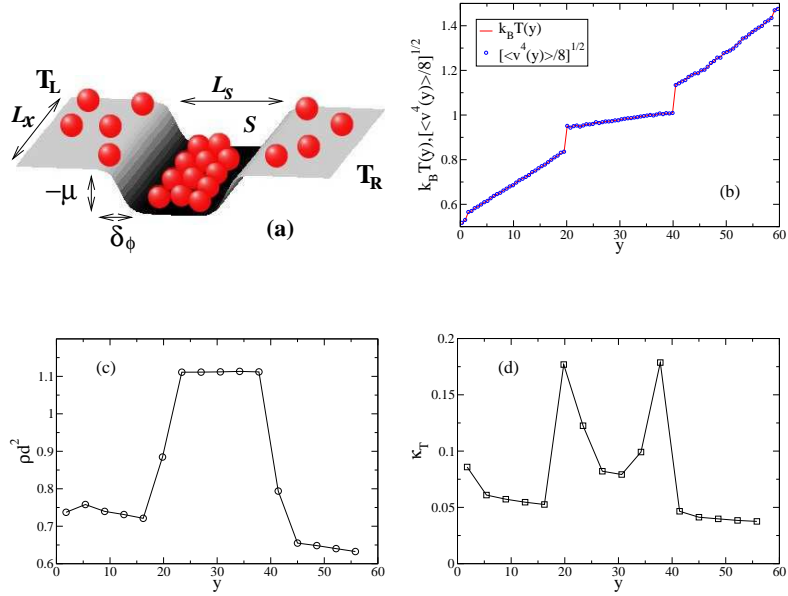}
\hskip .2cm
\includegraphics[width=5.cm]{mu13-tprof.eps}
\vskip 1cm
\includegraphics[width=5.cm]{dens-prof.eps}
\hskip .2cm
\includegraphics[width=5.cm]{comprs-prof.eps}
\end{center}
\caption{(Color online)  
(a) A schematic diagram of the system showing the liquid and 
solid regions produced by the external chemical potential of depth $-\mu$.
The various dimensions mentioned in the text are also marked in the figure.
(b) Plot of temperature profile $\kb T(y)$ and fourth moment of velocity
$m\sqrt{\langle v^4(y) \rangle/8}$ at $\mu=13$.
(c) The local density profile along $y$-direction at $\mu=13$.
(d) The isothermal compressibility $\kappa_T$ as a function of $y$ at $\mu=13$. Compressibility
shows strong peaks near the interfaces. Due to small size, interfacial
enhancement of compressibility permeates right through the whole of solid
region. In (c) and (d) lines are guides to eye.
}
\label{sys}
\end{figure}

We consider a two dimensional (2d) system of $N$ atoms of average density 
$\r = N/A$ within a rectangular box of area $A=L_x\times L_y$ 
(Fig.\ref{sys}(a)). The applied potential, $\phi(\rv)=-\mu$ for 
$\rv\in S$ which goes to zero with a hyperbolic tangent profile of
width $\d_\phi$ elsewhere,
enhances the density ($\r_s>\r$) in the central region
$S$ of area $A_s=L_x\times L_s$, ultimately leading it to a solid like 
phase. $T_L$ and $T_R > T_L$ are the temperatures of the two heat reservoirs 
in contact with the liquid regions at either end.

A large number of recent studies in lower
dimensions has shown that heat conductivity is, in fact, divergent 
as a function of system size\cite{lippi,bonet,grass}. Thus it is more
sensible to directly calculate the heat current $j_E\,(\,>0\,)$ flowing 
from the high to low temperature, or the conductance 
 of the system $G = j_E/\D T$ (or resistance $R=1/G$),  $\D T \,(\,>0\,)$ being the temperature difference, rather than the heat conductivity.

We report results for $1200$ particles interacting via the soft disk potential
$u(r_{ij})=1/r_{ij}^{12}$ taken within an area of $24\times 60$.
In absence of any external potential, a 2d system of soft disks
at this density $\r\approx0.83$ remains in the fluid phase.
The length and the energy scales are set by the soft disk diameter $d=1$,
and temperature $\kb T$ respectively while the time scale is set by
$\tau_s=\sqrt{m d^2/\kb T}$. The unit of energy flux $j_E$
is thus $(\kb T/\tau_s d)$. The unit of resistance and
conductance are $\tau_s d$ and $(\tau_s d)^{-1}$ respectively.
Periodic boundary conditions are applied in the $x$-direction.  
We use the standard velocity Verlet scheme of molecular dynamics 
(MD)\cite{frenkel} with equal time update of time-step $\d t$,
except when the particles collide with the `hard walled' heat reservoirs at 
$y=0$ and $y=L_y$. We treat the collision between the particles and the
reservoir as that between a hard disk of unit diameter colliding
against a hard, structure-less wall. If the time, $\tau_c$, of the next
collision with any of the two reservoirs at either end is smaller than
$\delta t$, the usual update time step of the MD simulation, we update
the system with $\tau_c$. During collisions with the walls Maxwell
boundary conditions are imposed to simulate the velocity of an atom
emerging out of a reservoir at temperatures ${T}_L$ (at $y=0$) or ${T}_R$
(at $y=L_y$)\cite{bonet}:

\bea
f(\vec v)=\frac{1}{\sqrt{2\pi}}\left(\frac{m}{k_B T_W}\right)^{3/2}
|v_y|\exp\left(-\frac{m {\vec v}^2}{2k_B T_W}\right) 
\eea
where $T_W$ is the temperature ($T_L$ or $T_R$) of the wall on which
the collision occurs. During each collision, energy is exchanged between 
the system and the bath. In the steady state, the average heat current 
flowing through the system can, therefore, be found easily by computing the
net heat loss from the system to the two baths  (say $Q_L$
and $Q_R$ respectively) during a large time interval $\tau$. 
The steady state heat current is given by 
$ \la J \ra
= \lim_{\tau \to \infty} Q_L/\tau = -\lim_{\tau \to \infty} Q_R /\tau$.    
In the steady state the heat current (the heat flux density integrated over
$x$) is independent of $y$. This is a requirement coming from current
conservation. For a homogeneous system $j_E=\la J \ra/L_x$.
However if the system has inhomogeneities then the flux
density itself can have a spatial dependence and in general we can
have $j_E=j_E(x,y)$. In our simulations we have looked at $j_E(x,0)$ 
and $j_E(x,L_y)$. 

\section{Results}
The system is first allowed to reach the steady state in a temperature 
gradient with the two walls at right and left being maintained at 
temperatures of $k_B T_R = 1.5$ and $k_B T_L = 0.5$ such that  
the current density integrated over the whole $x$-range is
the same at all $y$.  The local temperature can be defined as
$\kb T(y) = \langle 1/2~mv^2(y)\rangle$, 
where the averaging is done locally over strips of width $d=1$
and length $L_x$. 
If local thermal equilibrium (LTE) is maintained we should have, 
$\langle v^4(y) \rangle = 8(\kb T(y)/m)^2$. 
We find $\kb T(y)$ and $\langle v^4(y) \rangle$ as a function of
distance $y$ from cold to hot reservoir (Fig.~\ref{sys}(b)). 
From Fig.~\ref{sys}(b)
it is evident that the temperature profile is almost linear in the
single phase regions, with sharp increase
near the interfaces and LTE is approximately valid in all regions.
With increased $\mu$,
the temperature difference between the edges of the solid region decreases
indicating an enhancement of heat conductance within the solid.
The temparature jumps at the interfaces is a measure of the Kapitza or contact
resistance ($R_K$)\cite{kapitza} defined as,
\begin{eqnarray}
R_K = \frac{\Delta {T}}{j_E}
\end{eqnarray}
where $\Delta {T}$ is the difference in temperature across the
interface. The Kapitza resistance increases with increasing
trapping potential. 
It is evident that the interfaces are the regions of the
highest resistance in the system. This large resitance can be traced
back to large density mismatch at the contact of two phases. 
In Fig.\ref{sys}(c) we plot the local density profile $\r(y)d^2$. The
trapping region shows large density corresponding to the solid. Also the
colder liquid near the reservoir on the left shows a larger density than the hotter
liquid near the one on the right. In  Fig.\ref{sys}(d) we plot the local
compressibility $\kappa_T(y)$ defined via $\kappa_T=\r^{-2}(\p\r/\p\mu)_T$. 
Surprisingly, the compressibility of the interfaces
is very large making the narrow solid region
also unusually compressible pointing to the presence of 
large local number fluctuation. 
\begin{figure}[t]
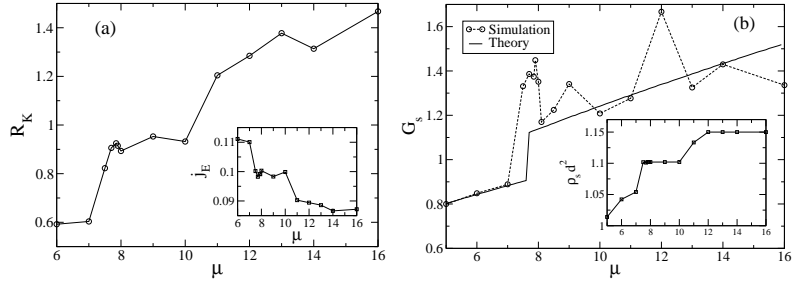

\begin{center}
\includegraphics[width=5.cm]{rk-j-epl.eps}
\hskip .2cm
\includegraphics[width=5.2cm]{epl-Conduct.eps}
\end{center}
\caption{(left)(a)Plot of the Kapitza resistance, $R_K$, expressed in units of 
$\tau_s d$ as a function of the potential depth $\mu$, shows a jump at the layering transition. 
(right)(b) Plot of the thermal conductance of the solid region, $G_s$
[in units of $(\tau_s d)^{-1}$] as a function of $\mu$. The points
denote simulation data and the solid line a free volume type calculation
of heat conductance $[G_s]_{fv}$.
The inset shows the corresponding change in solid density $\r_s d^2$.
}
\label{kapr}
\end{figure}

In Fig.~\ref{kapr}(a) we have plotted the Kapitza
resistance $R_k$ across the solid liquid interface, averaged over the
two interfaces, as a function of the
strength of the external potential $\mu$. The inset of Fig.~\ref{kapr}(a)
shows the heat flux through the system as a function of
$\mu$. As $\mu$ increases, the atoms from the surrounding liquid get
attracted into the potential well and the density of the liquid
becomes lower. The density mismatch at the solid-liquid
interface therefore increases progressively.
This figure shows fairly sharp increase in $R_k$ as well as sharp
decrease in current density $j_E$ near $\mu=8$ and $12$. As we will see later,
these are the $\mu$ values at which the solid undergoes two types
of layering transitions.

\begin{figure}[t]
\begin{center}
\includegraphics[width=5.cm]{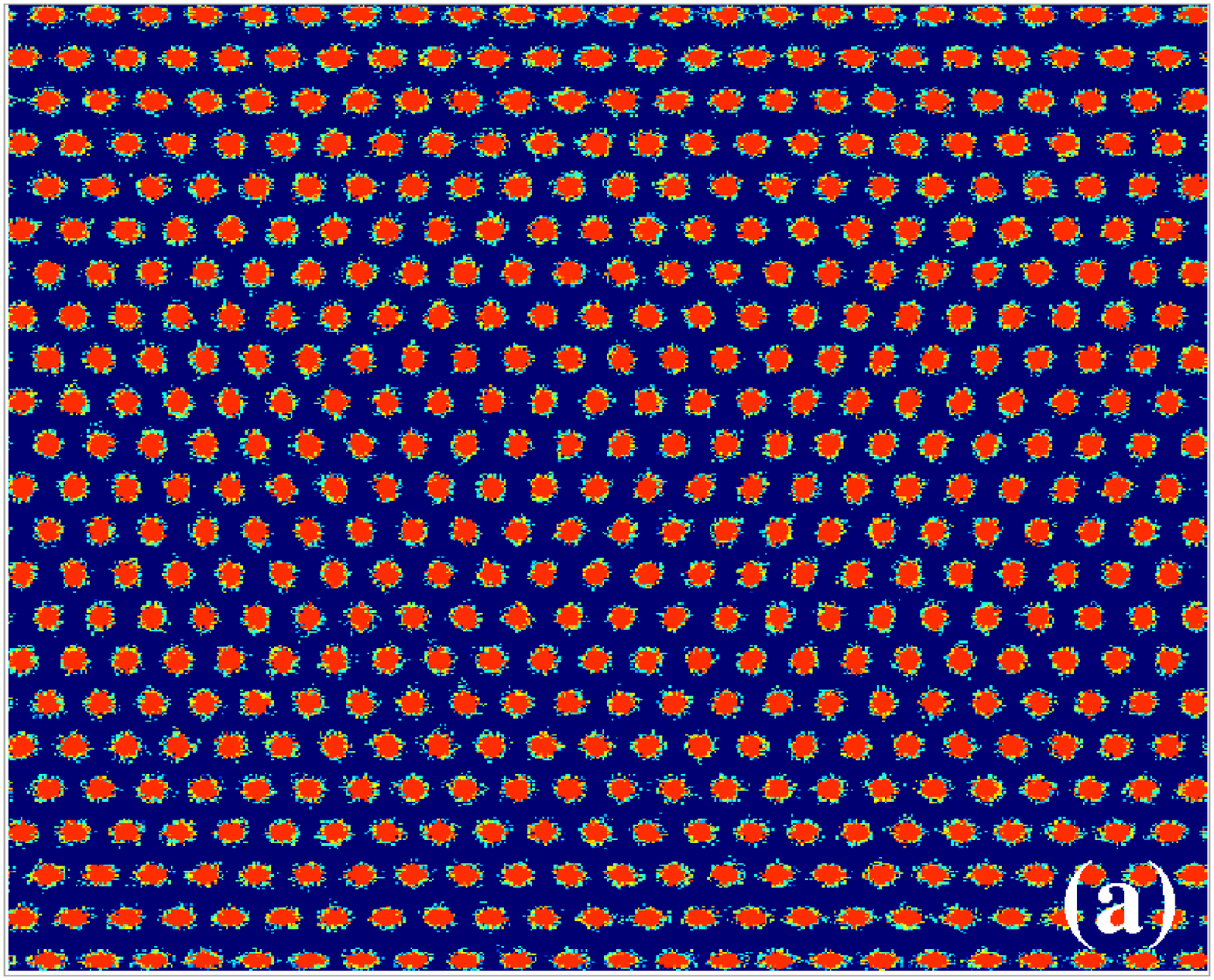}
\hskip .1cm
\includegraphics[width=5.cm]{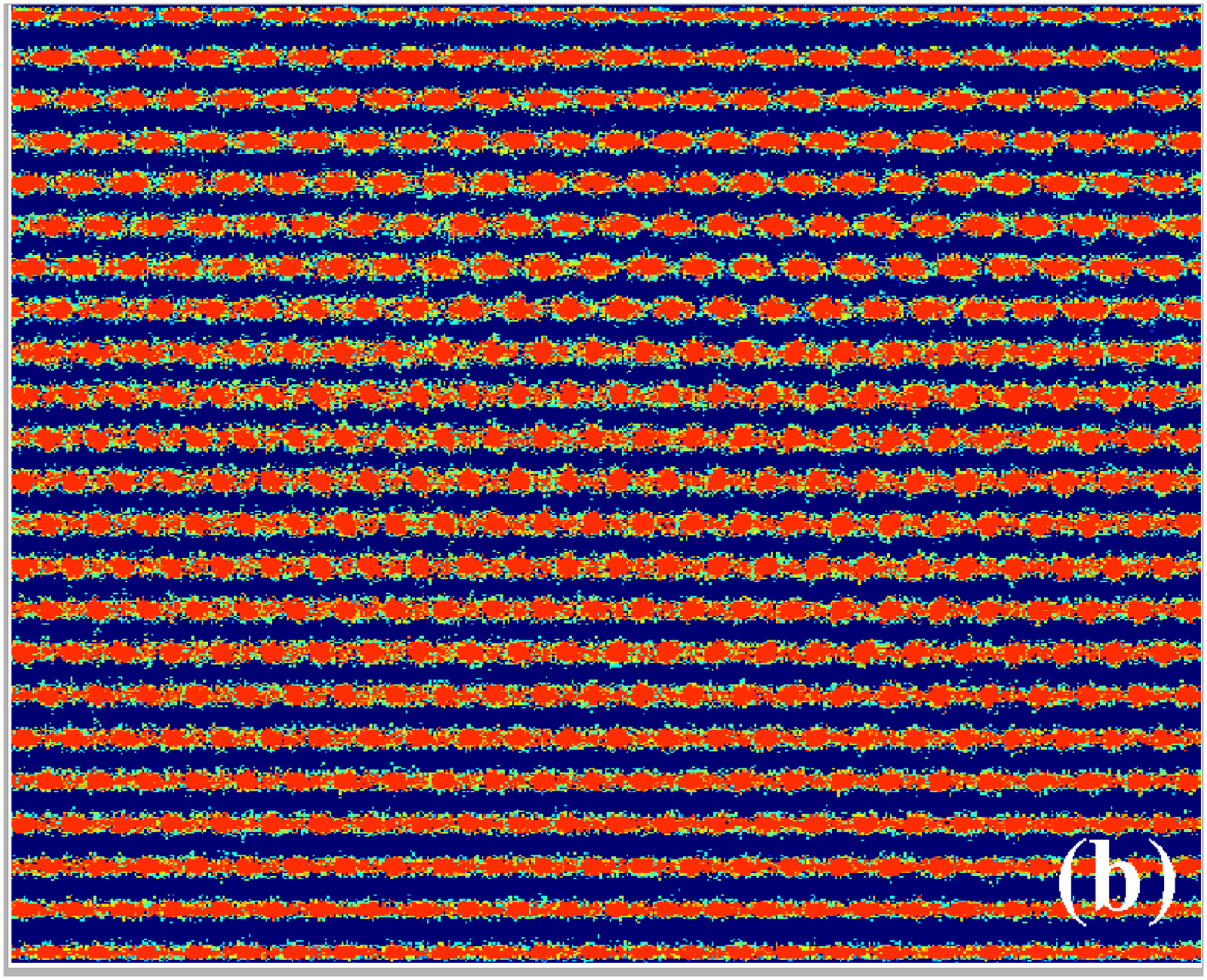}
\vskip .1cm
\includegraphics[width=5.cm]{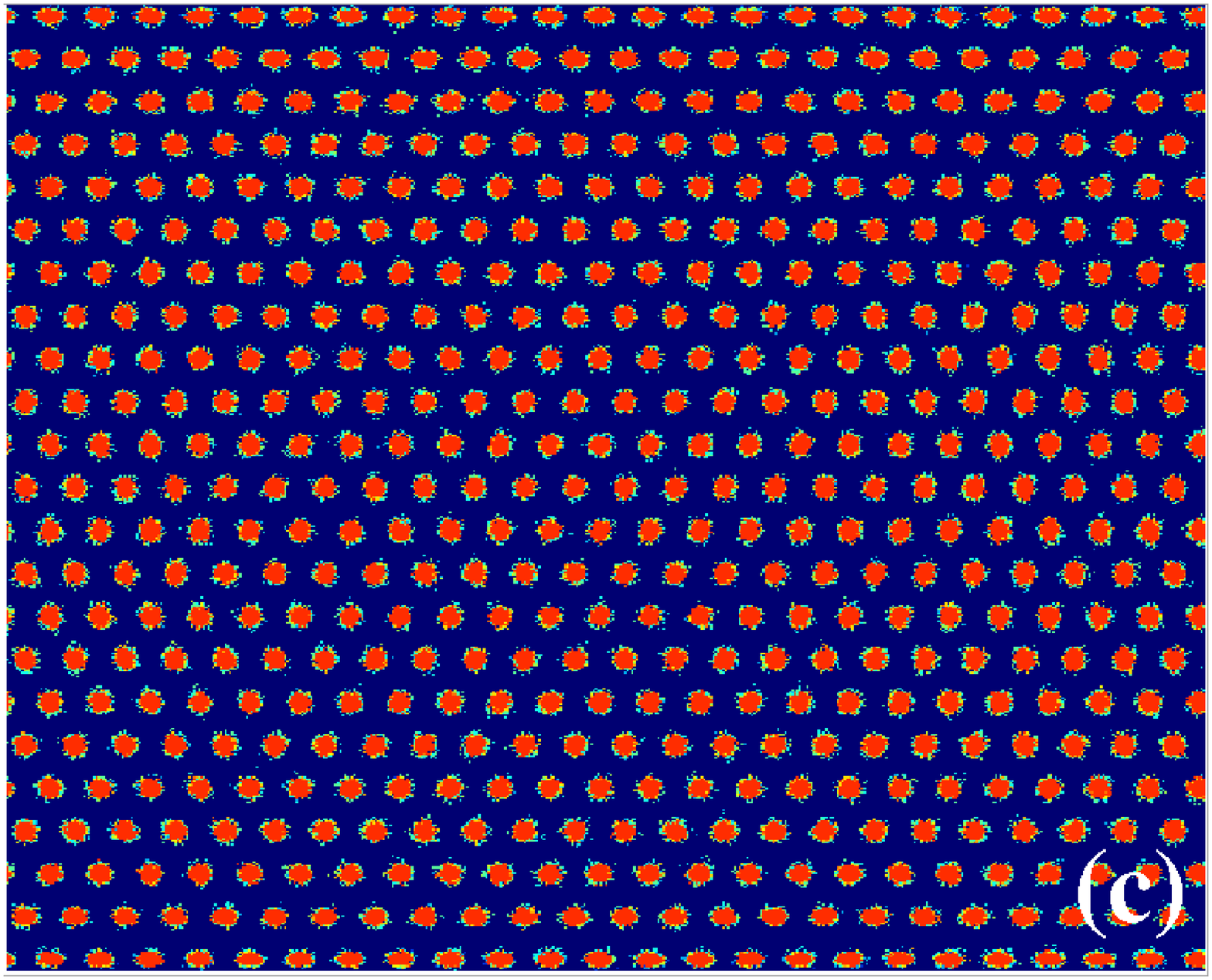}
\hskip .1cm
\includegraphics[width=5.cm]{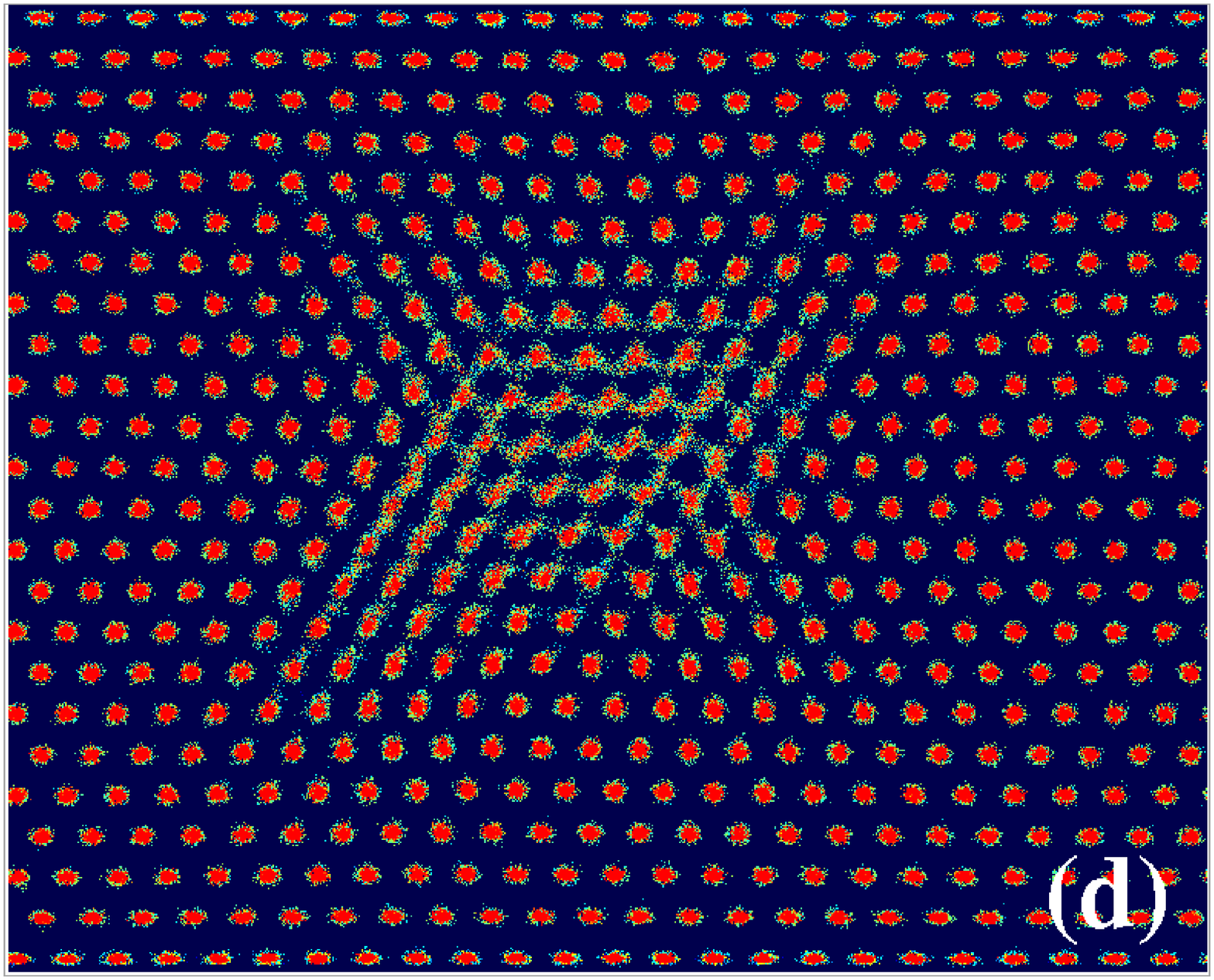}
\end{center}
\caption{(Color online)
Overlapped density plot of $500$ configurations in the region trapped
by external potential $\mu$: 
(a) A $23\times 23$ triangular lattice solid formed from a 
$23\times 22$ triangular lattice 
as the potential is increased from $\mu=7$ to $\mu=8$.
(b) Local density peaks hop in $x$-direction to incorporate $>23$ particles
in lattice planes in response to increased potential $\mu=11$.
(c) A $24\times 23$ triangular lattice solid at $\mu=12$. Notice the 
increase in particle numbers in the lattice planes.
(d) Configuration  obtained after $1.5\times 10^4 \d t$ as a $24\times 23$  
steady state solid at $\mu=16$ is quenched to $\mu=24$. This shows a
dislocation pair -- a $23$-layered region trapped in between a $24$-layered
solid. At steady state (after a time $10^5 \d t$) dislocations annihilate
to produce a $24\times 24$ triangular lattice solid.
Color code: blue (dark): low density and red (light): high density.
}
\label{dislo}
\end{figure}

In Fig.\ref{kapr}(b) we show the heat conductance in the solid region $G_s$ as
a function of strength of the trapping potential $\mu$. The inset in 
 Fig.\ref{kapr}(b) shows the change in the averaged density of the solid 
region $\r_s d^2$. The thick solid line
in   Fig.\ref{kapr}(b) is an analytical estimate obtained from a free volume
type calculation\cite{my-htcond} to be discussed in the next section.
The $\r_s d^2-\mu$ plot shows clear staircase-like sharp increases near
the same values of $\mu$ ($\approx$ 8 and 12) where sharp changes in thermal 
conductance occur. 
With increase in the strength of the trapping potential,
we observe two modes of
density enhancement: (A) A whole layer of particles enter to increase the
number of lattice planes in $y$-direction. This happens, e.g., as $\mu$ is
increased from $7$ to $8$. Thus in this mode the separation of lattice 
planes parallel to the liquid-solid interface
decreases (see Fig.\ref{dislo}(a)).  
(B)  
Each of the lattice planes parallel to the interface
grow by an atom thereby decreasing the interatomic separation within
each lattice plane.
This happens, e.g., as one increases $\mu$ from $10$ to $12$
(see Fig.\ref{dislo}(c)).
These two modes of density fluctuations
leave their signatures by enhancing the heat conductance $G_s$ - the 
effect of A being more pronounced than that of B. 

With increase in $\mu$, these two modes alternate
one after another, allowing the system to release
extra stress developed due to particle inclusion in one direction in one cycle,
by inclusion of particles in the perpendicular direction in the 
next cycle. Finally, at large enough $\mu$
the density of the solid region saturates ending the cycles. 
It is also interesting to observe, how the particles accomodate 
themselves going from Fig.\ref{dislo}(a)-(d). 
We find strained
triangular solids with $23\times23$, $24\times23$ and 
$24\times24$ unit cells at $\mu=8, 12$ and $24$ respectively 
(See Fig.\ref{dislo}). In the intermediate configurations one observes 
metastable dislocation pairs (Fig\ref{dislo}(d))
and peaks in the local particle density which 
corresponds to a few particles rapidly oscillating between two neighbouring
positions (Fig\ref{dislo}(b)) in order to maintain commensurability. Such 
rapid, localized, particle fluctuations may be observable in experiments.  

The layering transition in the solid by process (A) occurs via 
metastable dislocation formation and  
annihilation by incorporating  particles from the liquid
region. The kinematics of dislocation generation, transport and decay is 
controlled by diffusion which is a very slow 
process\cite{myfail} in a solid compared to particle collision and 
kinetic energy 
transfer times. Thus it is possible for a system with metastable 
dislocation pairs to reach an effective thermal steady state.  
Fig.\ref{dislo}.(d) shows overlapped configurations of the solid region
containing a dislocation-antidislocation pair, as the system is  
quenched from $\mu=16$ to $\mu=24$. The overlapped configurations are
separated by time $100\,\d t$ and collected after a time of 
$1.5\times 10^4\,\d t$ after the quench. At this stage the system is in a 
metastable state though, at the same time, maintaining LTE that we  check by 
computing $\la v^4(y)\ra$ and $\kb T(y)$ locally. This gives a heat conductance 
$G_s=2.29\, (\tau_s d)^{-1}$. After a further wait for $10^5 \,\d t$ 
the dislocations get annihilated. At this stage the whole solid region is 
transformed into  an equilibrium $24\times 24$-triangula lattice. 
On measuring the heat conductance now, we obtain
$G_s= 3.53\, (\tau_s d)^{-1}$. Thus with the annihilation of a single 
dislocation pair the conductance of the solid rises by about $54\%$!
Metastable configurations with dislocation pairs, therefore, have strikingly 
different thermal properties in this small system. Note that in the present 
system, configurations containing dislocations are always metastable since 
dislocations are either annihilated or are lost at the interface\cite{abhi}.

\noindent
\section{Free volume heat conductance}
Finally, we provide a brief sketch of an approximate theoretical approach 
for calculating heat conductance within the solid region. A detailed 
treatment of this approach is available in Ref.\cite{my-htcond}. 
The continuity of the energy density can be
utilized to obtain 
an exact expression for $\a$-th component of the heat flux 
density 
\bea
j_\a(\br) &=& j_\a^K(\br)+j_\a^U(\br) \nn\\
&=&\sum_i \d (\br - \br_i) h_i \bv_i^\a 
+ \f{1}{2} \sum_{i,j \neq i} \theta (x_i^\a-x^\a)
\prod_{\nu  \neq \a} \d (x^\nu-x_i^\nu)f_{ij}^\be( v_i^\be + v_j^\be).
\label{jr}
\eea
Here $\theta(x)$ is the Heaviside step function, $\d(\dots)$ is a
Dirac delta function, 
$h_i=m {\bf v}_i^2/2+\phi({\bf r}_i)+\sum_{i>j} u(r_{ij})$,
$\phi({\bf r}_i)$ is an onsite potential and $u(r_{ij})$ is 
inter-particle interaction.
The first term in Eq.\ref{jr}, $j_\a^K(\br)$, denotes 
the amount of energy carried by particle flux (convection)
and $j_\a^U(\br)$ denotes the net rate at
which work is done by particles on the left of $x^\a$ on the particles
on the right (conduction).
The $\a$-th component of the integrated heat current density over the 
solid region is,
\bea
\la I_\a \ra = 
 \sum_i \la~ h_i v_i^\a \ra -\f{1}{4} \sum_{i ,j \neq
  i} \left\la~ \f{\p u(r_{ij})}{\p r_{ij}}
\f{x^\a_{ij} x^{\be}_{ij}}{r_{ij}} (v_i^\be + v_j^\be)~ \right\ra.  
\label{totI}
\eea

In this study we focus on the average heat current density along $y$-direction,
$j_E = \la I_y \ra/L_x L_s$. We assume LTE and ignore conductance inside
solid~\cite{my-htcond}. Then assuming the conductance of our present
system to be simply proportional to that of a hard disk system with an 
effective diameter $\s$, the  heat conductance in units of $(\tau_s d)^{-1}$
can be expressed as,
\bea
G_s = \f{j_E}{\D T} = 
\left[3 \f{\r_s}{L_s} \f{y_c^2}{\tau_c}\right] \left(\f{d}{\s}\right)^2
\eea
where $\r_s$ is the average density of the solid, 
$y_c$ is the average separation between the colliding particles in $y$-direction
and $\tau_c$ is the mean collision time. 
The extra factor of $(d/\s)^2$ is due to the mapping of the soft disks
of diameter $d$ to effective hard disks of diameter $\s$.

We estimate $y_c^2$ and $\tau_c$ 
from the fixed neighbor free volume theory as
in Ref.\cite{my-htcond}.
Briefly, we assume that a (hard) test particle moves in the fixed
cage formed by the average positions of its neighbors and obtain the 
average values $[y_c^2]_{fv}$ from geometry and the timescale
$[\tau_c]_{fv}=c \sqrt{V_{fv}/\kb T}$ where $V_{fv}$ is the available
free volume of the test particle moving with a velocity derived from
the temperature $\kb T$. The effective hard disk diameter $\s$ and 
the constant $c$ of ${\cal O}(1)$ are both treated as fitting parameters. 
Using $\kb T=1$,  $d/\s=1.13$ and $c=0.4$ 
we obtain a fit to the $G_s-\mu$ curve with a  
layering transition from $22$ to $23$ layers near $\mu=8$. 
The fitted result, depicted as 
the solid line in Fig\ref{kapr}(b), is seen 
to reproduce most of the qualitative features of the simulation results, 
especially the jump in conductance due to the layering transition. 

\noindent 
\section{Conclusion}
We have shown that details of the structure
have a measurable effect on the thermal properties of the
trapped solid lying in contact with its liquid. 
In this study, we were
particularly interested in exploring the impact of structural changes,
viz. the layering transitions, on heat transport. One must, however, remember 
that, the layering transition is a finite size effect\cite{abhi} and
gets progressively less sharp as one goes to very large channel widths. 
An important consequence of this study is the
possibility that the thermal resistance of interfaces may be altered
using external potential which cause layering transitions in a trapped
nano solid. 
We have shown that metastable dislocations drastically reduce the conductance
of an otherwise defect free nano sized solid. 
Recently, electrical
\cite{my-econd} and thermal \cite{my-htcond} transport studies on confined
solid strips have also revealed strong signatures of such structural transitions
due to imposed external strain.
We believe that these phenomena have the potential for
useful applications e.g. as tunable thermal switches or in other nano
engineered devices.

\ack
We would like to thank Madan Rao, Abhishek Dhar, Tamoghna K. Das, 
Kurt Binder, Andrea Ricci and Peter Nielaba for discussions
and Martin Zapotocky for a critical reading of the manuscript. 
This work was partially supported by 
the Department of Science and Technology and CSIR (India).  


\end{document}